\documentclass[
    ,final            
  ]
  {aipproc}

\layoutstyle{6x9}

\usepackage{amsmath}

\begin{document}

\title{Effect of non-equilibrium term in two-particle correlation function on
electron-phonon collision integrals}


\author{A.Ya. Shul'man}{
address={Institute of Radio Engineering and Electronics of the RAS,
125009 Moscow, Russia.\\E-mail: ash@cplire.ru}
}

\begin{abstract}
The derivation of kinetic equation for one-particle distribution function 
$F(p,r,t)$ from BBGKY chain leads to the collision integral expressed in terms 
of the two-particle correlation function. The latter is in turn expressed by 
$F$ using Boltzmann's Stosszahlansatz that implies the neglect of initial 
one-time correlation function $g2(p1,p2,r,t)$ which is non-diagonal in the 
momentum space. In non-equilibrium case, however, it has been established that 
pair collisions generate the non-diagonal two-particle correlations that give 
the essential contribution to the current fluctuations of hot electrons. These 
correlations have been shown giving also a contribution to the collision 
integrals, i.e., to kinetic properties of nonequilibrium gas The expression for 
electron energy loss rate $P$ via phonons is re-derived in detail from this 
point of view. The order of value of the new term in $P$ in the cases of 
acoustic- and optical-phonon scattering is evaluated in the electron 
temperature approximation using the experimental study of hot-electron energy 
relaxation time in $n-InSb$ at helium temperatures.
\end{abstract}

\maketitle

\section{Introduction}
It was established that pair collisions generate in non-equilibrium case
non-zero two-particle correlations that are non-diagonal in momentum space and
give the essential contribution to the current fluctuations of hot electrons
\cite{GGK69-FTT,ASH70-FTT}. These correlations have been shown giving also a
contribution to the collision integrals, i.e., to kinetic properties of
non-equilibrium gas \cite{ASH02-cm,ASH03-Book}. The expression for electron
energy loss rate $P$ via phonons is re-derived from this point of view. The
formula for $P$ in terms of electron density-density correlation function
$\left\langle \rho_{k}\rho_{k}^{+}\right\rangle _{\omega}$ was obtained by
Kogan \cite{ShMK62-FTT} using Fermi Golden Rule. However, in order to come to
the well-known result for $P$ the fluctuation-dissipation theorem was
implicated to express $\left\langle \rho_{k}\rho_{k}^{+}\right\rangle
_{\omega}$ in terms of the dielectric function of the electron gas. On this
way the non-equilibrium correction has been lost. Here the contribution of the
non-diagonal part of the non-equilibrium pair correlator to phonon-electron
collision integral and to $P$ is obtained and explicitly calculated in the
electron temperature approximation. The contribution obtained does not contain
the extra powers of the small parameter of kinetic theory unlike the virial
expansions in equilibrium thermodynamics. The order of value of the new term
in $P$ is evaluated using the experimental study of hot-electron energy
relaxation time in $n-InSb$ at helium temperatures \cite{LOSh66-pss}.

\section{Basic designation}
Hamiltonian \newline $H=H_{e}+H_{p}+V_{ep}$\newline Electron part
\newline $H_{e}=\sum\limits_{\mathbf{k}}\hbar\Omega_{\mathbf{k}}a_{\mathbf{k}%
}^{+}a_{\mathbf{k}}+V_{ee}$, \newline where $V_{ee}$ stands for the Coulomb
electron-electron interaction\newline Phonon part \newline $H_{p}%
=\sum\limits_{\mathbf{f}}\hbar\omega_{\mathbf{f}}b_{\mathbf{f}}^{+}%
b_{\mathbf{f}}$ \newline and electron-phonon interaction\newline $V_{ep}%
=\frac{1}{\sqrt{V_{0}}}\sum\limits_{\mathbf{f}}\left(  c_{f}\rho_{\mathbf{f}%
}^{+}b_{\mathbf{f}}+c_{f}^{\ast}\rho_{\mathbf{f}}b_{\mathbf{f}}^{+}\right)  $
,\newline where $V_{0}$ is the normalizing volume.

Fourier components of the electron density operator\newline $\rho_{\mathbf{f}%
}=\sum\limits_{k}a_{\mathbf{k}+\frac{\mathbf{f}}{2}}^{+}a_{\mathbf{k}%
-\frac{\mathbf{f}}{2}},\,\,\,\,\,\,\,\,\,\,\,\rho_{\mathbf{f}}^{+}%
=\rho_{-\mathbf{f}}$

Phonon distribution function\newline $N_{\mathbf{f}}(t)=\left\langle \left(
b_{\mathbf{f}}^{+}b_{\mathbf{f}}\right)  _{t}\right\rangle $

Electron distribution function\newline $F_{\mathbf{k}}(t)=\left\langle \left(
a_{\mathbf{k}}^{+}a_{\mathbf{k}}\right)  _{t}\right\rangle $

The angle brackets denote the averaging with the density matrix corresponding
to the noninteracting electrons and phonons at initial instant of time. The
subscript $t$ indicates the Heisenberg representation for electron and 
phonon operators.

\section{Kinetic equation for phonons}
We obtain the kinetic equation for slow variables $N_{f}(t)$ , smoothing
Heisenberg equations of motion over mesoscopic time interval. With accuracy to
the second-order terms in $e-p$ interaction we have%
\[
\frac{\Delta N_{\mathbf{f}}(t)}{\Delta t}=-\frac{1}{\hbar^{2}\Delta
t}\left\langle S^{+}(t)\int\limits_{t}^{t+\Delta t}dt\prime\int\limits_{t}%
^{t\prime}dt\prime\prime\left[  V_{ep}^{0}(t\prime\prime),\left[  V_{ep}%
^{0}(t\prime),(b_{\mathbf{f}}^{+}b_{\mathbf{f}})_{t}\right]  \right]
S(t)\right\rangle .
\]
The superscript $0$ denotes operators in the interaction representation with
respect to the electron-phonon interaction $V_{ep}$ .

Calculations of the internal commutator lead to the desired result involving
the electron density-density correlation function:
\begin{equation}
\frac{dN_{\mathbf{f}}}{dt}=\frac{\left|  c_{f}\right|  ^{2}}{V_{0}\hbar^{2}%
}\left\{  \left\langle \rho_{\mathbf{f}}^{+}\rho_{\mathbf{f}}\right\rangle
_{-\omega_{f}}^{0}\left(  N_{\mathbf{f}}+1\right)  -\left\langle
\rho_{\mathbf{f}}\rho_{\mathbf{f}}^{+}\right\rangle _{\omega_{f}}%
^{0}N_{\mathbf{f}}\right\}  , \label{KinEq-ro+ro}%
\end{equation}
where the condition $\omega_{\mathbf{f}}\Delta t>>1$ was used. The function%
\[
\left\langle \rho_{\mathbf{f}}\rho_{\mathbf{f}}^{+}\right\rangle _{\omega
}=\int\limits_{-\infty}^{\infty}d\tau\left\langle \rho_{\mathbf{f}}%
(t+\tau)\rho_{\mathbf{f}}^{+}(t)\right\rangle \exp\left(  i\omega\tau\right)
\]
is the spectral density of the correlation function of the electron density fluctuations.

\subsection{Spectral density of the high-frequency electron
fluctuations in the electron temperature approximations}
The spectral density must be calculated without $V_{ep}$ taken into account.
From equations of motion for the electron operators we obtain in the random
phase approximation:
\begin{equation}
\left\langle \rho_{\mathbf{f}}^{+}\rho_{\mathbf{f}}\right\rangle _{\omega}%
^{0}=\frac{\left\langle \rho_{\mathbf{f}}^{+}\rho_{\mathbf{f}}\right\rangle
_{\omega}^{00}}{\left|  \epsilon\left(  \omega,\mathbf{f}\right)  \right|
^{2}},\label{ro+ro}%
\end{equation}
where $\epsilon(\omega,\mathbf{f})$ is the relative dielectric function of
the electron gas embedded into the dielectric medium and 
$\left\langle \rho_{\mathbf{f}}^{+}\rho_{\mathbf{f}}\right\rangle_{\omega}^{00}$
is the spectral density of the density-density correlation
function for non-interacting electrons. The latter is connected with one-time
correlation function $\left\langle \rho_{\mathbf{f}}(\mathbf{k})\rho
_{\mathbf{f}}^{+}(\mathbf{k}_{1})\right\rangle _{t}$ by equations of motion
for free electrons. The one-time correlation function is defined as%
\[
\left\langle \rho_{\mathbf{f}}(\mathbf{k})\rho_{\mathbf{f}}^{+}(\mathbf{k}%
_{1})\right\rangle _{t}\equiv\left\langle a_{\mathbf{k}-\frac{\mathbf{f}}{2}%
}^{+}a_{\mathbf{k}+\frac{\mathbf{f}}{2}}a_{\mathbf{k}_{1}+\frac{\mathbf{f}}%
{2}}^{+}a_{\mathbf{k}_{1}-\frac{\mathbf{f}}{2}}\right\rangle _{t}.
\]
The one-time correlation function is independent of time (in steady state)
under following condition on the electron energies $\Omega_{k}=\hbar
k^{2}/2m:$%
\[
\Omega_{\mathbf{k}-\frac{\mathbf{f}}{2}}-\Omega_{\mathbf{k+}\frac{\mathbf{f}%
}{2}}+\Omega_{\mathbf{k}_{1}+\frac{\mathbf{f}}{2}}-\Omega_{\mathbf{k}%
_{1}-\frac{\mathbf{f}}{2}}\equiv\frac{\hbar}{m}\mathbf{f}\left(
\mathbf{k}_{1}-\mathbf{k}\right)  =0.
\]
Because $\mathbf{f}\neq0$ then it is necessary to be $\mathbf{k}%
_{1}=\mathbf{k}$ . Therefore, this condition provides the one-time correlator
to be the integral of motion during the time interval $\Delta t$ like the
distribution functions $N_{\mathbf{f}}$ and $F_{\mathbf{k}}$.

\subsection{Kinetic equation with account for the contribution of
the non-equilibrium pair correlations}
The substitution of the explicit expression Eq.(\ref{ro+ro})for correlators to
the kinetic equation Eq.(\ref{KinEq-ro+ro}) gives rise to%

\begin{align}
\frac{dN_{\mathbf{f}}}{dt}  &  =\frac{2\left|  C_{f}\right|  ^{2}}{V_{0}%
\hbar^{2}\left|  \epsilon\left(  \omega_{\mathbf{f}},\mathbf{f}\right)
\right|  ^{2}}\sum\limits_{\mathbf{k}}\left\langle a_{\mathbf{k}%
+\frac{\mathbf{f}}{2}}^{+}a_{\mathbf{k}-\frac{\mathbf{f}}{2}}a_{\mathbf{k}%
-\frac{\mathbf{f}}{2}}^{+}a_{\mathbf{k}+\frac{\mathbf{f}}{2}}\right\rangle
_{t}\delta\left(  \Omega_{\mathbf{k}+\frac{\mathbf{f}}{2}}-\Omega
_{\mathbf{k}-\frac{\mathbf{f}}{2}}-\omega_{\mathbf{f}}\right)  \left(
N_{\mathbf{f}}+1\right)  -\label{KinEq-a+a}\\
&  -\frac{2\left|  C_{f}\right|  ^{2}}{V_{0}\hbar^{2}\left|  \epsilon\left(
\omega_{\mathbf{f}},\mathbf{f}\right)  \right|  ^{2}}\sum\limits_{\mathbf{k}%
}\left\langle a_{\mathbf{k}-\frac{\mathbf{f}}{2}}^{+}a_{\mathbf{k}%
+\frac{\mathbf{f}}{2}}a_{\mathbf{k}+\frac{\mathbf{f}}{2}}^{+}a_{\mathbf{k}%
-\frac{\mathbf{f}}{2}}\right\rangle _{t}\delta\left(  \Omega_{\mathbf{k}%
+\frac{\mathbf{f}}{2}}-\Omega_{\mathbf{k}-\frac{\mathbf{f}}{2}}-\omega
_{\mathbf{f}}\right)  N_{\mathbf{f}}.\nonumber
\end{align}

Taking into account the commutation rules for $a_{\mathbf{k}}^{+}%
a_{\mathbf{k}_{1}}$ it is easy to obtain%
\[
\left\langle a_{\mathbf{k}}^{+}a_{\mathbf{k}_{1}}a_{\mathbf{k}_{1}}%
^{+}a_{\mathbf{k}}\right\rangle _{t}=F_{\mathbf{k}}(t)-\left\langle
a_{\mathbf{k}}^{+}a_{\mathbf{k}_{1}}^{+}a_{\mathbf{k}_{1}}a_{\mathbf{k}%
}\right\rangle _{t}.
\]
According to the results from [1]%
\[
\left\langle a_{\mathbf{k}}^{+}a_{\mathbf{k}_{1}}^{+}a_{\mathbf{k}_{1}%
}a_{\mathbf{k}}\right\rangle =F_{\mathbf{k}}F_{\mathbf{k}_{1}}+\phi\left(
\mathbf{k},\mathbf{k}_{1}\right)  ,
\]
where $\phi\left(  \mathbf{k},\mathbf{k}_{1}\right)  $ is a non-equilibrium
correction to the pair correlation function.

Finally, the kinetic equation for phonons with non-equilibrium corrections of
electron pair correlators takes the form:%
\begin{align}
\frac{dN_{\mathbf{f}}}{dt}  &  =\frac{2\left|  C_{f}\right|  ^{2}}{V_{0}%
\hbar^{2}\left|  \epsilon\left(  \omega_{f},\mathbf{f}\right)  \right|  ^{2}%
} \nonumber\\
& \times\sum\limits_{\mathbf{k}}\left[
F_{\mathbf{k}+\frac{\mathbf{f}}{2}}\left(
1-F_{\mathbf{k}-\frac{\mathbf{f}}{2}}\right)  -\phi\left(  \mathbf{k}%
+\frac{\mathbf{f}}{2},\mathbf{k}-\frac{\mathbf{f}}{2}\right)  \right]  \left(  N_{\mathbf{f}%
}+1\right)  \delta\left(  \Omega_{\mathbf{k}+\frac{\mathbf{f}}{2}}%
-\Omega_{\mathbf{k}-\frac{\mathbf{f}}{2}}-\omega_{\mathbf{f}}\right)
-\label{KinEqu}\\
& -
\left[
F_{\mathbf{k}-\frac{\mathbf{f}}{2}}\left(  1-F_{\mathbf{k}+\frac{\mathbf{f}%
}{2}}\right)  -\phi\left(  \mathbf{k}-\mathbf{f}/2,\mathbf{k}+\mathbf{f}%
/2\right)  \right]  N_{\mathbf{f}}\delta\left(  \Omega_{\mathbf{k}%
+\frac{\mathbf{f}}{2}}-\Omega_{\mathbf{k}-\frac{\mathbf{f}}{2}}-\omega
_{\mathbf{f}}\right) \nonumber
\end{align}
Certainly, it needs to keep in mind the Eq.(\ref{KinEqu}) describes the part
of full time dependence of the phonon occupation numbers that resulting from
electron-phonon interaction only.

It should be stressed that for ordinary classical equilibrium
electron gas there are no pair correlations between different
states of momentum space. This fact can be easily seen from the
form of the Gibbs distribution. In conclusion of these
considerations it needs to underline that the contribution of
non-equilibrium correlations to kinetics should be expected in the
cases when the particles considered are scattered by
non-equilibrium many-particle system.

\section{Electron energy loss rate via phonons}
The energy loss rate of electrons due to phonons per unit volume can be
obtained using the relation%
\[
P=\frac{1}{V_{0}}\sum\limits_{\mathbf{f}}\hbar\omega_{\mathbf{f}}%
\frac{dN_{\mathbf{f}}}{dt}.
\]
It is seen from the kinetic equation for phonons Eq.(\ref{KinEqu}) that the
loss rate can be broken up into two parts:
\[
P=P_{1}+P_{2},
\]
where the first term on the right hand side is related to the one-particle
electron distribution function and it is described by the known expressions.
The second one is due to the correlation contribution determined by the
function $\phi$ .We assume during the calculations of $P$ the
electron-electron collisions are sufficiently frequent in order to provide the
use of the electron temperature approximation for hot electrons. As a result
of these assumptions the following formulae may be derived
\begin{align}
P_{1}(T,T_{0}) &  =\sum_{\mathbf{f}}\frac{2\left|  C_{f}\right|  ^{2}}%
{(V_{0}\hbar)^{2}\left|  \epsilon(\omega_{f},\mathbf{f})\right|  ^{2}}%
\hbar\omega_{f}\left[  N_{\mathbf{f}}^{0}(T)-N_{\mathbf{f}}\right]
\label{P1}\\
&  \times\sum_{\mathbf{k}}\left[  F_{\mathbf{k}}^{0}(T)-F_{\mathbf{k+f}}%
^{0}(T)\right]  \delta\left(  \Omega_{\mathbf{k+f}}-\Omega_{\mathbf{k}}%
-\omega_{\mathbf{f}}\right)  \nonumber
\end{align}
and
\begin{equation}
P_{2}(T,T_{0})=-\sum\limits_{\mathbf{f}}\frac{2\left|  C_{f}\right|  ^{2}%
}{(V_{0}\hbar)^{2}\left|  \epsilon(\omega_{\mathbf{f}},\mathbf{f}\right|
^{2}}\hbar\omega_{\mathbf{f}}\sum\limits_{\mathbf{k}}\phi\left(
\mathbf{k}+\mathbf{f},\mathbf{k}\right)  \delta\left(  \Omega_{\mathbf{k}%
+\mathbf{f}}-\Omega_{\mathbf{k}}-\omega_{\mathbf{f}}\right)  \label{P2}%
\end{equation}
Here $N_{\mathbf{f}}^{0}(T)=\left[  \exp\left(  \hbar\omega_{\mathbf{f}%
}/T\right)  -1\right]  ^{-1}$ is the equilibrium Bose-Einstein distribution
with electron temperature $T$ and $N_{\mathbf{f}}$ is the true non-equilibrium
phonon distribution function. Further, we neglect possible phonon heating and
consider that the phonons remain in thermal equilibrium at temperature $T_{0}%
$, so $N_{\mathbf{f}}=N_{\mathbf{f}}^{0}\left(  T_{0}\right)  $.

The function $\phi\left(  \mathbf{k},\mathbf{k}_{1}\right)  $ was found in
\cite{ASH70-FTT} under accepted assumptions for the case of nondegenerate
electron gas and can be presented in the form:
\begin{equation}
\phi(\mathbf{k},\mathbf{k}_{1})=Q(T,T_{0})\frac{\partial F_{\mathbf{k}}^{0}%
}{\partial T}\frac{\partial F_{\mathbf{k}_{1}}^{0}}{\partial T} \label{phiT}%
\end{equation}
where
\[
Q(T,T_{0})=\frac{T^{2}N_{c}(T)}{nc_{e}}\left[  \frac{P(T,T_{0})/\left(
T-T_{0}\right)  }{dP/dT}-1\right]  .
\]
Here $N_{c}(T)$ and $nc_{e}=3n/2$ are the effective density of states and
specific heat capacitance of electron gas with concentration $n$,
respectively. The expression for $\phi$ in \cite{ASH70-FTT} was obtained
without accounting for the contribution of $\phi$ itself in $P$ .Therefore,
the quantity $P_{1}$ was only implied as the total energy loss rate. Now
evidently we have the differential equation for determination of
$P_{2}(T,T_{0})$ or full $P$ since this quantity is defined by the right hand
side of the equality Eq.(\ref{P2}) including $P_{2}(T,T_{0})$ itself. The
situation is quite resembling the self-consistent field approach in plasma
theory. It needs to stress that the substitution of the total loss rate $P$
instead of $P_{1}$ in the expression for $\phi\left(  k,k_{1}\right)  $ is
justified because all symmetry properties of the electron-electron and
electron-phonon integrals of collisions used in the derivation are not
destroyed by introducing $\phi\left(  \mathbf{k},\mathbf{k}_{1}\right)  $ in
collision integrals.

It is of interest to estimate the interrelation between $P_{1}$
and $P_{2}$. Instead carrying out the integrations in Eqs.
(\ref{P1})-(\ref{P2}) and following solution of a nonlinear
differential equation for $P$ (or $P_{2}$) we can use the
experimental data from Ref. \cite{LOSh66-pss} where hot-electron
energy relaxation time at helium temperatures in $n-InSb$ was
investigated and all necessary information on $n$, $P$, $dP/dT$,
and $T$ can be extracted from. It was shown there that at low
heating the electron energy loss is due to acoustic phonon
scattering and next it is changed by longitudinal optical (LO)
phonon scattering. So we can evaluate $P_{2}/P_{1}$ for both types
of the electron-phonon scattering.

It is necessary to carry out preliminary the integrations over $\mathbf{k}$ in
Eqs.(\ref{P1})-(\ref{P2}) for all essential parameters be explicitly
introduced. Taking into account nondegeneracy of electron gas the expressions
for $P_{1}$and $P_{2}$ can be transform as following
\begin{align}
P_{1}(T,T_{0})  &  =(T-T_{0})\frac{2ms^{2}}{(2\pi)^{2}\hbar^{3}}%
\sum_{\mathbf{f}}\frac{2\left|  C_{f}\right|  ^{2}}{\hbar V_{0}\left|
\epsilon(\omega_{f},\mathbf{f})\right|  ^{2}}F_{\mathbf{f/}2}^{0}%
(T),\label{P1-ac}\\
P_{2}(T,T_{0})  &  =\pm\frac{9Tm^{2}s}{4c_{e}\hbar^{3}}\left[  \frac
{P(T,T_{0})/(T-T_{0})}{dP/dT}-1\right]  \sum_{\mathbf{f}}\frac{2\left|
C_{f}\right|  ^{2}}{\hbar V_{0}\left|  \epsilon(\omega_{f},\mathbf{f})\right|
^{2}}F_{\mathbf{f/}2}^{0}(T/2)\left[  1+O(T)\right] \label{P2ac}\\
F_{\mathbf{k}}^{0}(T)  &  =\frac{n}{N_{c}(T)}\exp\left(  -\frac{\hbar^{2}%
k^{2}}{2mT}\right)  .\nonumber
\end{align}
for acoustic phonon scattering. Here $s$ is the sound velocity and
it has been accounted for the near-elastic scattering of electrons
by acoustic phonons. While the sign of expression (\ref{P2ac}) for
$P_{2}$ leaves indeterminate since implicit $O(T)$ terms may
redefine it, the order of magnitude of $P_{2}/P_{1}$ can be found.
The expression Eq.(\ref{P1-ac}) is just the Kogan's formula from
\cite{ShMK62-FTT} corrected for the screening of electron-phonon
interaction.

In the case of LO phonon scattering we obtain
\begin{align}
P_{1}(T,T_{0})  &  =\frac{m^{2}T\hbar\omega_{LO}}{\pi^{2}\hbar^{4}}\frac
{n}{N_{c}(T)}\left(  1-\mathsf{e}^{-\hbar\omega_{LO}/T}\right)  \left[
N_{\mathbf{f}}^{0}(T)-N_{\mathbf{f}}^{0}(T_{0})\right] \nonumber\\
&  \times\sum_{\mathbf{f}}\frac{2\left|  C_{f}\right|  ^{2}}{\hbar
V_{0}\left|  \epsilon(\omega_{f},\mathbf{f})\right|  ^{2}f}\exp(-\frac
{\hbar^{2}}{2mT}\left[  \frac{f^{2}-f_{LO}^{2}}{2f}\right]  ^{2})
\label{P1-LO}%
\end{align}%
\begin{align}
P_{2}(T,T_{0})  &  =\pm\frac{m^{2}\hbar^{2}\omega_{LO}^{2}}{2(2\pi)^{2}%
\hbar^{4}}\frac{n}{N_{c}(T)}\left[  \frac{P(T,T_{0})/(T-T_{0})}{dP/dT}%
-1\right]  \exp\left(  -\frac{\hbar\omega_{LO}}{T}\right) \nonumber\\
&  \times\sum_{\mathbf{f}}\frac{2\left|  C_{f}\right|  ^{2}}{\hbar
V_{0}\left|  \epsilon(\omega_{f},\mathbf{f})\right|  ^{2}f}\exp(-\frac
{2\hbar^{2}}{2mT}\left[  \frac{f^{2}-f_{LO}^{2}}{2f}\right]^{2})\left[
1+O(T)\right]  \label{P2-LO}%
\end{align}
where $f_{LO}^{2}=2m\omega_{LO}/\hbar$ .

The sums over $\mathbf{f}$ in corresponding expressions are of the same order
of magnitude. Therefore, we have to evaluate following ratios
\begin{equation}
\frac{P_{2}}{P_{1}}=\frac{3\pi^{2}T}{T-T_{0}}\left[  \frac{P(T,T_{0}%
)/(T-T_{0})}{dP/dT}-1\right]  \label{Ratio-ac}%
\end{equation}
for the acoustic phonon scattering and
\begin{equation}
\frac{P_{2}}{P_{1}}=\frac{\hbar\omega_{LO}}{8T}\left[  \frac{P(T,T_{0}%
)/(T-T_{0})}{dP/dT}-1\right]  \label{Ratio-LO}%
\end{equation}
for the low-temperature LO phonon scattering  ($\hbar\omega_{LO}/T>>1$). 
Turning to experimental data
\cite{LOSh66-pss} and using relation between $dP/dT$ and measured electron
energy relaxation time $\tau_{P}=nc_{e}/(dP/dT)$ we obtain
for the acoustic scattering $P_{2}/P_{1}%
\simeq1.3$ at $T \simeq7.8K$, $P \simeq6\cdot10^{-3}W/cm^{3}$,
$\tau_{P} \simeq4\cdot10^{-7}s$ and
$P_{2}/P_{1}\simeq0.6$ at $T \simeq15.8K$, $P%
\simeq3.8 \cdot10^{-2}W/cm^{3}$, $\tau_{P}\simeq0.8\cdot10^{-7}s$
for LO phonon scattering. All data are taken at $T_{0}=4.2K$.
These estimations may be considered as an indication that the pair
correlations contribute to hot electron energy loss via phonons
comparable with common accepted expressions.
\subsection{Concluding remarks}
It is of importance to note in conclusion that the expression for
$\phi\left( k,k_{1}\right)$ found in the limit of high frequency
of the electron-electron collisions does contain coupling
constants of neither electron-electron nor electron-phonon
interactions. Therefore, the corresponding corrections to the
integral of phonon-electron collisions do not contain additional
powers of the small parameters of the kinetic theory. The
interesting but still open question: are there analogous
contributions to the hydrodynamic equations?
\begin{theacknowledgments}
This work was partially supported by Russian Foundation for Basic
Researches.
\end{theacknowledgments}

\end{document}